\documentclass[aps,pre,reprint,showpacs]{revtex4-1}

\usepackage{graphics,epsfig}
\usepackage{amsmath,amssymb}
\usepackage{bm}

\begin{document}
\title{Scaling of alloy interfacial properties under compositional strain}
\author{Zhi-Feng Huang}
\affiliation{Department of Physics and Astronomy, Wayne State University, 
Detroit, Michigan 48201, USA}
\date{\today}

\begin{abstract}
Complex morphologies and microstructures that emerge during materials growth 
and solidification are often determined by both equilibrium and kinetic 
properties of the interface and their crystalline anisotropies. However 
limited knowledge is available for the alloying and particularly the 
compositionally generated elastic effects on these interface characteristics. 
Here we systematically investigate such compositional effects on the interfacial 
properties of an alloy model system based on the phase-field-crystal analysis, 
including the solid-liquid interfacial free energy, 
kinetic coefficient, and lattice pinning strength. Scaling relations for 
these interfacial quantities over various ranges of material parameters 
are identified and predicted. Our results indicate the important effects 
of couplings among mesoscopic and microscopic length scales of alloy 
structure and concentration, and also the influence of compressive and 
tensile interface stresses induced by composition variations.
The approach developed here provides an efficient way to systematically 
identify these key material properties beyond the traditional atomistic 
and continuum methods.
\end{abstract}


\maketitle

\section{Introduction}

Properties of surfaces and interfaces are among the vital 
factors controlling material crystallization and microstructural dynamics. 
Typical examples include the crucial effects of the liquid-solid interfacial 
free energy on dendritic solidification \cite{Haxhimali06}, 
eutectic or peritectic growth \cite{re:asta09,Ghosh15}, and the evolution of 
film surface nanostructures such as quantum dots \cite{re:aqua13} or nanowires 
\cite{Tsivion11}. Properties governing system kinetics, in particular the 
interface mobility or kinetic coefficient (defined as the ratio between 
interface velocity and undercooling or supersaturation), also significantly affect 
the material microstructures and morphologies during e.g., crystal nucleation, 
ordering, and dendrite formation \cite{re:asta09,Tang13,re:granasy04}.

Despite both fundamental and technological importance of these interfacial
properties in the characterization, understanding and modeling of materials
growth, it remains a great challenge to experimentally or computationally 
determine their accurate values, anisotropies, and particularly their variations 
with material parameters and growth or processing conditions. Significant 
difficulty exists in the corresponding experimental measurements, with 
limited data available for the interfacial energy anisotropy of alloys
\cite{Liu01,*Napolitano04,Niederberger06} and the interface kinetic 
coefficient of only few pure metals \cite{Glicksman67,Rodway91}.
Most calculations rely on atomistic simulations via molecular dynamics 
(MD) and Monte Carlo (MC) methods \cite{re:asta09,Hoyt01,Becker07,
Amini08,Kerrache08,Tang13,Wilson15},
or continuum approaches based on phase field \cite{Wheeler93},
Ginzburg-Landau \cite{Shih87,Wu15}, or classical density functional
\cite{re:harrowell86,re:mikheev91} theory. However, it is 
computationally challenging to conduct any systematic studies across 
a reasonable range of material parameters, particularly for alloy systems 
for which very limited results are available to date. For example, the 
alloying effect on the kinetic coefficient is not yet understood, with only 
few data obtained from recent MD simulations of binary ordered phases which 
estimated the value of kinetic coefficient at the
melting temperature \cite{Kerrache08,Tang13,Wilson15}. 
Most MD and MC studies of alloy solid-liquid interfacial energy 
have been focused on either zero \cite{Becker07} or a specific 
finite value \cite{Amini08} of atomic size difference between alloy 
components, while a systematic understanding of the effects of 
the associated compositional strain \cite{Larche85,re:guyer95}, 
which is known to play an important role on determining material 
microstructures, properties and growth morphologies, is still lacking.

The focus of this work is on identifying the key factors governing 
alloy crystal-melt interfacial properties, particularly the effects
generated by compositional stresses and by the couplings among
mesoscopic structural amplitudes and alloy concentration and the 
underlying microscopic crystalline lattice. This leads to new 
scaling behaviors of alloy interfacial free energy $\gamma$ and 
kinetic coefficient $\mu_k$, a reversal of $\gamma$ anisotropy 
caused by compositional strain, and an interface lattice pinning 
effect that is crucial in determining system growth mechanisms 
and dynamics. Our findings reveal that these results not only 
depend on the interface orientation as expected, but also on the
impacts of interface preferential segregation and the corresponding
compositionally induced interface stresses.

To obtain a generic understanding of such effects, here we adopt a model 
alloy system which incorporates the crystalline symmetry from a simple 
but fundamental aspect. It also enables us to systematically
examine the varying conditions of compositional strain.
More specifically, we develop a new nonadiabatic complex amplitude 
approach for binary alloys based on the phase field crystal (PFC) method 
\cite{re:elder02,*re:elder07,re:huang08,*re:huang10,emmerich12}.
In PFC models lattice symmetry is built in the system free energy 
functional via the selection and competition between different modes 
of characteristic microscopic length scales
(e.g., minimum 1 mode for two-dimensional (2D) triangular and 
three-dimensional (3D) bcc structures \cite{re:elder02,*re:elder07}, 
2 modes for fcc and hcp \cite{wu10,re:greenwood10,*re:greenwood11b}, 
and 3 modes for simple cubic \cite{re:greenwood10,*re:greenwood11b} 
and also some complex 2D phases and superlattices \cite{Mkhonta13}). 
In this work we focus on the 2D triangular system, to emphasize 
the fundamental aspects of the alloying effects and the essential
features of our approach which can be readily generalized to other systems.
In addition, the properties identified here can be used for the study
of various 2D crystallization phenomena such as the epitaxial growth of 
sub-monolayer islands for metallic alloy overlayers \cite{Einax13}
or novel 2D materials \cite{vanderZande13}.

This paper is organized as follows. In Sec. \ref{sec:ampl} the complex
amplitude formulation for binary alloy system is derived and presented,
showing new results of nonadiabatic corrections that originate from the 
coupling between microscopic and mesoscopic spatial scales. The corresponding
interface equations of motion and the analytic expressions of interfacial
quantities derived are given in Sec. \ref{sec:interface_eqs}. In 
Sec. \ref{sec:properties} detailed numerical calculations of liquid-solid
interfacial properties are conducted, with new scaling behaviors and
effects of compositionally generated stresses being identified and discussed.
A brief summary of our results is given in Sec. \ref{sec:summary}.

\section{Nonadiabatic amplitude equations for binary PFC}
\label{sec:ampl}

We start from the PFC model equations governing the dynamics of a 
dimensionless atomic density variation field $n=(\rho-\rho_l)/\rho_l$ 
and an alloy concentration field $\psi=(\rho_A-\rho_B)/\rho$ for a binary 
alloy system, where $\rho=\rho_A+\rho_B$ is the total atomic number density, 
$\rho_{A(B)}$ is the density of A(B) atoms, and $\rho_l$ is a reference state 
density. These dynamic equations can be written in a rescaled form \cite{re:huang10b}
\begin{eqnarray}
&&\partial n / \partial t = \nabla^2 \frac{\delta F}{\delta n} 
+ m \nabla^2 \frac{\delta F}{\delta \psi} + {\bm \nabla} \cdot {\bm \eta}_n, 
\label{eq:bpfc_n}\\
&&\partial \psi / \partial t = m \nabla^2 \frac{\delta F}{\delta n} 
+ \nabla^2 \frac{\delta F}{\delta \psi} + {\bm \nabla} \cdot {\bm \eta}_{\psi},
\label{eq:bpfc_psi}
\end{eqnarray}
where the mobility contrast $m=(M_A-M_B)/(M_A+M_B)$ with $M_A$ ($M_B$) the 
atomic mobility of alloy component A (B), ${\bm \eta}_n$ and ${\bm \eta}_{\psi}$ 
are noise fields, and for one-mode PFC the free energy functional is given by
\begin{eqnarray}
&F = & \int d{\bm r} \left \{ - \frac{1}{2} \epsilon n^2 
+ \frac{1}{2} n \left ( \nabla^2 + q_0^2 \right )^2 n
+ \frac{1}{3} g_2 n^3 + \frac{1}{4} n^4 \right. \nonumber\\
&& + \frac{1}{2} K_0 \left | {\bm \nabla} \psi \right |^2 
+ \frac{1}{2} (w_0 + 2v_1 n + g n^2) \psi^2
+ \frac{1}{4} u_0 \psi^4 \nonumber\\
&& \left. + 2\alpha n \left ( \nabla^2 + \nabla^4 \right ) (n \psi) \right \}.
\label{eq:F_pfc}
\end{eqnarray}
Here $\epsilon$ is proportional to the temperature distance from the melting 
point, $q_0=1$ after rescaling over a length scale of lattice spacing, and 
$g_2$, $K_0$, $w_0$, $v_1$, $g$, $u_0$ are phenomenological model parameters 
determining system properties including elastic moduli and the phase diagram 
(e.g., eutectic or isomorphous; see Ref. \cite{re:huang10b} for more detailed 
description). Also $\alpha$ is the solute expansion coefficient \cite{Larche85}
defined as $\alpha=\partial \ln a / \partial \psi$ (with $a$ the alloy lattice 
constant) which characterizes the atomic size mismatch between alloy components. 
It gives rise to the compositional strain $\alpha \psi$ generated by local 
composition variations \cite{Larche85,re:guyer95}.

In the standard amplitude formulation of a crystalline system 
\cite{re:goldenfeld05,*re:athreya06,re:elder10a,re:huang10b}, the 
density field $n$ is expanded as $n=n_0+\sum_j A_j e^{i {\bm q}_j^0 
\cdot {\bm r}} + {\rm c.c.}$, where $n_0$ is the average density variation
(i.e., $n_0=(\rho_0-\rho_l)/\rho_l$ with $\rho_0$ the average number density) 
and ${\bm q}_j^0$ are the basic wave vectors of the crystalline lattice 
($j=3$ for triangular structure with ${\bm q_1^0} = 
-q_0 ( \sqrt{3} \hat{x}/2 + \hat{y}/2 )$, ${\bm q_2^0} = q_0 \hat{y}$, 
and ${\bm q_3^0} = q_0 (\sqrt{3} \hat{x}/2 - \hat{y}/2 )$). 
Both the complex amplitudes $A_j$ and alloy concentration field $\psi$ are 
assumed to vary on ``slow'' scales that can be separated from the underlying 
``fast'' scales of crystalline lattice. However, for thin enough 
interfaces such an adiabatic approximation of scale separation is no 
longer valid, and nonadiabatic corrections \cite{Huang13} are needed 
to account for the coupling between mesoscopic (amplitudes and 
concentration) and microscopic (lattice) length scales.

\begin{figure}
\centerline{\includegraphics[clip,width=0.39\textwidth]{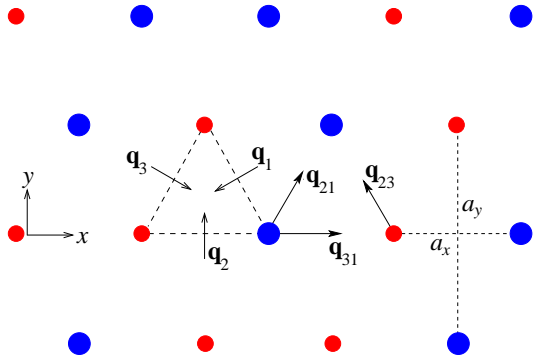}}
\caption{(Color online) Schematic of a triangular lattice for a binary 
A-B system. The directions of ${\bm q}_j = (1+\delta_0){\bm q}_j^0$ and
${\bm q}_{ij}={\bm q}_i-{\bm q}_j$ (with $i,j=1,2,3$) and lattice spacings
$a_x$, $a_y=\sqrt{3} a_x$ are indicated.}
\label{fig:lattice}
\end{figure}

To derive the corresponding nonadiabatic amplitude equations, we
follow the procedure of multiple-scale analysis outlined in 
Refs. \cite{re:huang10b,Huang13}, including:
(i) Separate ``slow'' vs ``fast'' scales of $(X=\epsilon^{1/2}x, 
Y=\epsilon^{1/2}y, T=\epsilon t)$ vs $(x,y,t)$, and assume concentration 
field $\psi=\psi(X,Y,T)$, complex amplitudes $A_j=A_j(X,Y,T)$, and
$n_0=n_0(X,Y,T)$ in the expansion of $n$ , 
(ii) conduct the multiple-scale expansion on the binary PFC equations 
(\ref{eq:bpfc_n}) and (\ref{eq:bpfc_psi}) 
and apply the solvability conditions, (iii) keep the nonadiabatic
coupling between ``slow'' and ``fast'' spatial scales across interfaces of
various orientations (e.g., 6 directions for the triangular structure 
shown in Fig. \ref{fig:lattice}), and (iv) given 
the bulk compositional elastic effect, rewrite amplitudes
$A_j=A_j' e^{i {\bm q}_j^0 \cdot {\bm u}_c}$ with displacement vector 
${\bm u}_c=\delta_0 {\bm r}$ to address the compositional strain in 
alloy systems, where $\delta_0=\sqrt{1-2\alpha\psi_s}-1$ (with $\psi_s$ 
the equilibrium composition in the solid bulk) can be identified from 
the corresponding free energy minimization \cite{re:huang10b}. 
In the case of triangular symmetry, to the lowest order we obtain
\begin{widetext}
\begin{eqnarray}
&\partial A_1' / \partial t =& -q_0^2 (1-m^2) \frac{\delta \mathcal{F}}{\delta A_1'^*}
- (1-m^2) \int_x^{x \pm a_x} \frac{dx'}{a_x} \int_y^{y+a_y} \frac{dy'}{a_y} \nonumber\\
&& \left [ f_{p_{11}} e^{i {\bm q}_1 \cdot {\bm r}'}
+ f_{p_1} e^{-i {\bm q}_2 \cdot {\bm r}'} + f_{p_0} e^{-i {\bm q}_3 \cdot {\bm r}'} 
+ f_{p_{12}} e^{i {\bm q}_{12} \cdot {\bm r}'} 
+ f_{p_{13}} e^{i {\bm q}_{13} \cdot {\bm r}'}
+ f_{p_{33}}^* e^{i {\bm q}_{23} \cdot {\bm r}'}
+ f_{p_2}^* e^{i {\bm q}_{32} \cdot {\bm r}'} \right ]
+ \eta_1, \label{eq:A1}\\
&\partial A_2' / \partial t =& -q_0^2 (1-m^2) \frac{\delta \mathcal{F}}{\delta A_2'^*}
- (1-m^2) \int_x^{x \pm a_x} \frac{dx'}{a_x} \int_y^{y+a_y} \frac{dy'}{a_y} \nonumber\\
&& \left [ f_{p_1}^* e^{-i {\bm q}_1 \cdot {\bm r}'}
+ f_{p_2} e^{i {\bm q}_2 \cdot {\bm r}'} + f_{p_3}^* e^{-i {\bm q}_3 \cdot {\bm r}'} 
+ f_{p_{21}} e^{i {\bm q}_{21} \cdot {\bm r}'} 
+ f_{p_{33}}^* e^{i {\bm q}_{13} \cdot {\bm r}'}
+ f_{p_{11}}^* e^{i {\bm q}_{31} \cdot {\bm r}'}
+ f_{p_{23}} e^{i {\bm q}_{23} \cdot {\bm r}'} \right ]
+ \eta_2, \label{eq:A2}\\
&\partial A_3' / \partial t =& -q_0^2 (1-m^2) \frac{\delta \mathcal{F}}{\delta A_3'^*}
- (1-m^2) \int_x^{x \pm a_x} \frac{dx'}{a_x} \int_y^{y+a_y} \frac{dy'}{a_y} \nonumber\\
&& \left [ f_{p_0}^* e^{-i {\bm q}_1 \cdot {\bm r}'}
+ f_{p_3} e^{-i {\bm q}_2 \cdot {\bm r}'} + f_{p_{33}} e^{i {\bm q}_3 \cdot {\bm r}'} 
+ f_{p_{31}} e^{i {\bm q}_{31} \cdot {\bm r}'}
+ f_{p_2}^* e^{i {\bm q}_{12} \cdot {\bm r}'}
+ f_{p_{11}}^* e^{i {\bm q}_{21} \cdot {\bm r}'}
+ f_{p_{32}} e^{i {\bm q}_{32} \cdot {\bm r}'} \right ]
+ \eta_3, \label{eq:A3}\\
&\partial n_0 / \partial t =& \nabla^2 \frac{\delta F}{\delta n_0}
+ m \nabla^2 \frac{\delta \mathcal{F}}{\delta \psi}
- \int_x^{x \pm a_x} \frac{dx'}{a_x} \int_y^{y+a_y} \frac{dy'}{a_y} \nonumber\\
&& \left [ \left ( f_{p_0} + mf_{p'_0} \right ) e^{i {\bm q}_{13} \cdot {\bm r}'}
+ \left ( f_{p_1} + mf_{p'_1} \right ) e^{i {\bm q}_{12} \cdot {\bm r}'} 
+ \left ( f_{p_3} + mf_{p'_3} \right ) e^{i {\bm q}_{32} \cdot {\bm r}'} + {\rm c.c.} \right ]
+ {\bm \nabla} \cdot {\bm \eta}_0, \label{eq:n0}\\
&\partial \psi / \partial t =& m \nabla^2 \frac{\delta F}{\delta n_0}
+ \nabla^2 \frac{\delta \mathcal{F}}{\delta \psi}
- \int_x^{x \pm a_x} \frac{dx'}{a_x} \int_y^{y+a_y} \frac{dy'}{a_y} \nonumber\\
&& \left [ \left ( mf_{p_0} + f_{p'_0} \right ) e^{i {\bm q}_{13} \cdot {\bm r}'}
+ \left ( mf_{p_1} + f_{p'_1} \right ) e^{i {\bm q}_{12} \cdot {\bm r}'}
+ \left ( mf_{p_3} + f_{p'_3} \right ) e^{i {\bm q}_{32} \cdot {\bm r}'} + {\rm c.c.} \right ]
+ {\bm \nabla} \cdot {\bm \eta}_{\psi_0}, \label{eq:psi}
\end{eqnarray}
\end{widetext}
where $a_x=2\pi/q_x$ and $a_y=4\pi/q_y$ are lattice spacings, 
$q_y=q_0(1+\delta_0)$, $q_x=\sqrt{3}q_y/2$, 
${\bm q}_j = (1+\delta_0){\bm q}_j^0$, ${\bm q}_{ij}={\bm q}_i-{\bm q}_j$, and
the integration terms are the nonadiabatic corrections 
representing the coupling between ``slow'' and ``fast'' length scales 
(i.e., meso-micro scale coupling) which is missing in previous amplitude 
analysis of alloy systems \cite{re:elder10a,re:huang10b}.
Coefficients $f_{p_{ij}}$, $f_{p_j}$ and $f_{p'_j}$ are functions of
slowly-varying amplitudes $A_j'$, $n_0$, and concentration field $\psi$, i.e.,
\begin{eqnarray}
&& f_{p_0} = 3q_0^2 \left [ (6n_0+2g_2) A_1'A_3'^* + 3 \left ( A_1'^2A_2'
+ A_2'^*{A_3'^*}^2 \right ) \right ], \nonumber\\
&& f_{p_1} = 3q_0^2 \left [ (6n_0+2g_2) A_1'A_2'^* + 3 \left ( A_1'^2A_3'
+ {A_2'^*}^2A_3'^* \right ) \right ], \nonumber\\
&& f_{p_2} = 4q_0^2 \left [ (3n_0+g_2) A_2'^2 + 6A_1'^*A_2'A_3'^* \right ], \nonumber\\
&& f_{p_3} = 3q_0^2 \left [ (6n_0+2g_2) A_2'^*A_3' + 3 \left ( A_1'A_3'^2
+ A_1'^*{A_2'^*}^2 \right ) \right ], \nonumber\\
&& f_{p_{11}} = 4q_0^2 \left [ (3n_0+g_2) A_1'^2 + 6A_1'A_2'^*A_3'^* \right ], \nonumber\\
&& f_{p_{33}} = 4q_0^2 \left [ (3n_0+g_2) A_3'^2 + 6A_1'^*A_2'^*A_3' \right ], \nonumber\\
&& f_{p_{jk}} = 21q_0^2 A_j'^2 A_k'^* ~~(j \neq k), \nonumber\\
&& f_{p'_0} = 6q_0^2 \left [ f_{\psi} A_1'A_3'^*
-q_0^2\alpha \left ( A_1' \mathcal{G}_3'^* A_3'^* + A_3'^* \mathcal{G}'_1 A_1' 
\right ) \right ], \nonumber\\
&& f_{p'_1} = 6q_0^2 \left [ f_{\psi} A_1'A_2'^*
-q_0^2\alpha \left ( A_1' \mathcal{G}_2'^* A_2'^* + A_2'^* \mathcal{G}'_1 A_1' 
\right ) \right ], \nonumber\\
&& f_{p'_3} = 6q_0^2 \left [ f_{\psi} A_3'A_2'^*
-q_0^2\alpha \left ( A_3' \mathcal{G}_2'^* A_2'^* + A_2'^* \mathcal{G}'_3 A_3' 
\right ) \right ], \label{eq:fp}
\end{eqnarray}
where 
\begin{eqnarray}
& f_{\psi} = g\psi + 2q_0^2\delta_1^0\alpha, \quad
\delta_1^0=-2q_0^2\alpha\psi_s,& \\
& \mathcal{G}'_{1,3} = \nabla^2 \mp 2iq_x\partial_x - iq_y\partial_y, \quad
\mathcal{G}'_2=\nabla^2+2iq_y\partial_y.& 
\end{eqnarray}
In the above amplitude equations the noise terms satisfy the conditions
(with $i,j=1,2,3$, $\mu, \nu = x,y$, 
$\vartheta_i=\vartheta_0=\vartheta_{\psi}=\vartheta=1/7$, $T$ the temperature, 
and $\Gamma$ a rescaled constant \cite{re:huang10b})
\begin{eqnarray}
& \langle \eta_j \rangle = \langle {\bm \eta}_0 \rangle 
= \langle {\bm \eta}_{\psi_0} \rangle = 0, & \nonumber\\
& \langle \eta_i \eta_j \rangle = \langle {\bm \eta}_0 \eta_j \rangle
= \langle {\bm \eta}_0 \eta_j^* \rangle = \langle {\bm \eta}_{\psi_0} \eta_j \rangle 
= \langle {\bm \eta}_{\psi_0} \eta_j^* \rangle = 0, & \nonumber\\
& \langle \eta_i \eta_j^* \rangle = 2(1-m^2) \vartheta_i q_0^2 \Gamma k_BT
\delta ({\bm r} - {\bm r'}) \delta (t-t') \delta_{ij}, & \nonumber\\
& \langle \eta_0^{\mu} \eta_0^{\nu} \rangle = 2 \vartheta_0 
\Gamma k_BT \delta ({\bm r} - {\bm r'}) \delta (t-t') \delta^{\mu\nu}, & \nonumber\\
& \langle \eta_{\psi_0}^{\mu} \eta_{\psi_0}^{\nu} \rangle = 2 \vartheta_{\psi} 
\Gamma k_BT \delta ({\bm r} - {\bm r'}) \delta (t-t') \delta^{\mu\nu}, & \nonumber\\
& \langle \eta_{\psi_0}^{\mu} \eta_0^{\nu} \rangle = 2m\vartheta_{\psi} 
\Gamma k_BT \delta ({\bm r} - {\bm r'}) \delta (t-t') \delta^{\mu\nu}. &
\end{eqnarray}

In Eqs. (\ref{eq:A1})--(\ref{eq:psi}) the free energy functional $\mathcal{F}$
is given by
\begin{eqnarray}
&\mathcal{F}&=\int d{\bm r} \left [ \sum\nolimits_j 
\left | \mathcal{G}'_jA_j' \right |^2 
+ \frac{1}{2} K_0 \left | {\bm \nabla} \psi \right |^2 \right. \label{eq:F}\\
&& \left. -\sum\nolimits_j \left ( 2q_0^2 \alpha \psi + \delta_1^0 \right ) 
\left (A_j'^* \mathcal{G}'_j A_j' + {\rm c.c.} \right ) + f(A_j',n_0,\psi) \right ],
\nonumber
\end{eqnarray}
where $f$ is the bulk free energy density, i.e., 
\begin{eqnarray}
f&=& \sum_j \left ( -\epsilon + 3n_0^2 + 2g_2n_0 + {\delta_1^0}^2 
+ 4q_0^2\delta_1^0\alpha\psi \right ) |A_j'|^2 \nonumber\\
&+& \frac{3}{2} \sum_j |A_j'|^4 
+ (6n_0+2g_2) \left ( \prod_j A_j' + {\rm c.c.} \right ) \label{eq:f}\\
&+& 6 \sum_{j<k} |A_j'|^2 |A_k'|^2 
+ \frac{1}{2} \left ( -\epsilon + q_0^4 \right ) n_0^2 
+ \frac{1}{3} g_2 n_0^3 + \frac{1}{4} n_0^4 \nonumber\\
&+& \left ( \frac{1}{2} w_0 + \frac{1}{2} gn_0^2 + v_1 n_0 
+ g \sum\nolimits_j |A_j'|^2 \right ) \psi^2 + \frac{1}{4} u_0 \psi^4. \nonumber
\end{eqnarray}
Note that the free energy functional given in Eq. (\ref{eq:F}) 
(and also Eq. (\ref{eq:F_pfc})) is invariant with respect to 
$\alpha \rightarrow -\alpha$ and $\psi \rightarrow -\psi$.
Also the free energy terms in Eqs. (\ref{eq:F}) and (\ref{eq:f}) 
incorporate the coupling between composition profile and system 
elasticity and also between mesoscopic structural amplitudes and 
alloy concentration (i.e., meso-meso scale coupling).

To identify the elastic energy of this alloy system, we rewrite
the amplitudes as
$A_j'=|A_j| e^{i{\bm q}_j^0 \cdot {\bm u}_s} \equiv 
\phi e^{i{\bm q}_j^0 \cdot {\bm u}_s}$
in the limit of small deformations, where ${\bm u}_s=(u_x^s,u_y^s)$ is the 
displacement field and $\phi=|A_j|$. Substituting it into 
Eq. (\ref{eq:F}) we obtain the system elastic energy (to the leading order)
\begin{widetext}
\begin{eqnarray}
&F_{\rm elastic} =& \int d{\bm r} \left \{ 3q_0^4 (1+\delta_0^2) \phi^2
\left [ \frac{3}{2} \left ( u_{xx}^2+u_{yy}^2 \right ) + u_{xx} u_{yy}
+ 2u_{xy}^2 \right ] + 3\delta_1^0 \left ( 4q_0^2\alpha\psi+\delta_1^0 \right )
\phi^2 \right. \nonumber\\
&& \left. -12 \left ( q_0^2\alpha\psi+\delta_1^0 \right ) \phi \nabla^2 \phi
+ 3q_0^2 \left ( 2q_0^2\alpha\psi+\delta_1^0 \right ) \phi^2 
\left [ 2(1+\delta_0) \left ( u_{xx} + u_{yy} \right ) 
+ u_{xx}^2+u_{yy}^2 + 2 \left ( u_{xy}^2 + \Omega_{xy}^2 \right )
\right ] \right \}, \label{eq:F_elastic}
\end{eqnarray}
\end{widetext}
where $u_{ij}=(\partial_i u_j^s + \partial_j u_i^s)/2$ is the linear strain tensor 
and $\Omega_{ij}=(\partial_i u_j^s - \partial_j u_i^s)/2$ is the rotation tensor
\cite{Larche85}. Terms in the 2nd line of Eq. (\ref{eq:F_elastic}) incorporate
the compositionally-induced interface elastic effects, given that both 
$\phi\nabla^2\phi$ and $2q_0^2\alpha\psi+\delta_1^0 = 2q_0^2\alpha(\psi-\psi_s)$ 
are nonzero only around the interface. These terms drive the process of solute
preferential segregation towards the liquid-solid interface and determine whether
the local elastic deformation is compressive or tensile (noting that around the
interface $\psi-\psi_s>0$ for $\psi_s<0$ but $\psi-\psi_s<0$ for $\psi_s>0$),
as will be further discussed in Sec. \ref{sec:properties}. 

\section{Interface equations of motion with scale coupling and pinning effects}
\label{sec:interface_eqs}

The above nonadiabatic amplitude equations can be further coarse-grained
to derive the interface equations of motion. We first apply the 
projection operator method \cite{re:elder01} around a fixed interface orientation
$\theta=\theta_0$ (the angle between the interface normal $\hat{\bm n}$ and 
the vertical $\hat{y}$ direction), so that different scales of variations can be 
separated for local curvilinear coordinates $u$ (along $\hat{\bm n}$) 
and $s$. The results are then extended to the general case of $\theta$ via 
a variation scheme \cite{Herring51,re:karma98} for the system free energy. 
Detailed procedure and the resulting interface equations are given in 
Appendix \ref{sec:dev_interface}. Here we focus on a simplified case
for which $n_0$ is assumed to be a constant due to the secondary 
effect of its variation in an alloy system. At the liquid-solid interface 
the anisotropic form of the generalized Gibbs-Thomson relation is then given by
\begin{equation}
\mu_k^{-1}(\theta,m) v_n  = -\Delta - \left ( \gamma + \gamma'' \right ) \kappa
-p_0(\theta) \sin(q h_n + \varphi) + \eta_v,
\label{eq:G-T}
\end{equation}
where $v_n$ is the normal velocity of the interface, $\kappa$ is the local 
curvature, $h_n$ represents the interface height, $q=|{\bm q}_j|$ or 
$|{\bm q}_{ij}|$, and $\eta_v$ is a noise term. 
The interface supersaturation is given by 
$\Delta = q_0^2 \Delta \psi_0 \delta\mu_{\psi}(u=0,s)$ (the same expression
as that of Ref. \cite{re:langer80} for isothermal solidification), 
where the miscibility gap 
$\Delta \psi_0 = \psi_0(+\infty) - \psi_0(-\infty) \equiv \psi_l - \psi_s$,
the chemical potential $\mu_{\psi}(u,s)=\delta \mathcal{F} / \delta \psi$,
and $\delta\mu_{\psi} = \mu_{\psi} - \mu_{\psi}^{\rm eq}$ with $\mu_{\psi}^{\rm eq}$
the equilibrium value determined by one-dimensional (1D) 
solutions $\psi_0(u)$ and $A_j^0(u)$ governing liquid-solid coexistence
(see Eqs. (\ref{eq:Ajpsi0})--(\ref{eq:psi0}) in Appendix \ref{sec:dev_interface}). 
The interface at $u=0$ is defined as a Gibbs surface satisfying the condition
$\int_{-\infty}^{+\infty} du [ \psi_0(u) - \psi_0(\pm \infty) ] = 0$.
In Eq. (\ref{eq:G-T}) the interfacial free energy is expressed as
\begin{eqnarray}
&& \gamma(\theta) = q_0^2 \int_{-\infty}^{+\infty} du \left \{ 
K_0 \left ( \partial_u \psi_0 \right )^2 + 4\sum\nolimits_j \left [ 
|\partial_u^2 A_j^0|^2 \right. \right. \nonumber\\
&& \left. + \left ( \beta_j^2/2 
+ 2q_0^2 \alpha \psi_0 + \delta_1^0 \right ) |\partial_u A_j^0|^2 
+ q_0^2 \alpha (\partial_u |A_j^0|^2)(\partial_u \psi_0) \right ] \nonumber\\
&& \left. + 2 \sum\nolimits_j (\partial_{\theta} \beta_j) 
\left [ i(\partial_u^2 A_j^0)(\partial_u {A_j^0}^*) + {\rm c.c.} \right ] \right \}, 
\label{eq:gamma}
\end{eqnarray}
with $\beta_{1,3} = \mp 2q_x \cos \theta + q_y \sin \theta 
= 2q_y \sin(\theta \mp \pi/3)$ and $\beta_2 = -2q_y \sin \theta$, 
and the kinetic coefficient $\mu_k(\theta,m)$ is determined by
\begin{equation}
\mu_k^{-1} = \int \frac{du}{1-m^2} \left \{ 
2\sum\nolimits_j |\partial_u A_j^0|^2 + q_0^2 \left [ \psi_0^2
- \psi_0^2(\pm \infty) \right ] \right \}.
\label{eq:mu_k}
\end{equation}

An important feature incorporated in Eq. (\ref{eq:G-T}) is the coupling to 
the underlying lattice structure, which results in a sine-Gordon type term 
$p_0 \sin(q h_n + \varphi)$ resembling a periodic pinning potential. The 
corresponding lattice pinning strength $p_0$ and phase $\varphi$ are
orientation-dependent, i.e.,
\begin{equation}
p_0 e^{i\varphi} = 2i \left [ p_A(\theta) + p_{\psi}(\theta) \right ],
\label{eq:p0}
\end{equation}
where $p_A$ originates from the meso-micro scaling coupling for amplitudes 
$A_j'$ given in Eqs. (\ref{eq:A1})--(\ref{eq:A3}), while $p_{\psi}$ originates 
from the scale coupling of $\psi$ field in Eq. (\ref{eq:psi}), 
with $p_{\psi}=0$ at $\theta=0$, $\pm \pi/3$ (${\bm q}_j$ directions)
and $p_{\psi} \neq 0$ at $\theta=\pi/2$, $\pm \pi/6$ (${\bm q}_{ij}$ 
orientations). Specifically,
\begin{eqnarray}
&p_{\psi} = q_0^2& \left \{ \left [ \int_0^{+\infty} du \psi_0(u) \int_u^{+\infty} du'
\right. \right. \nonumber\\
&& \left. - \int_{-\infty}^0 du \psi_0(u) \int_{-\infty}^u du' \right ] I(u') 
\label{eq:p_psi}\\
&& \left. - \left ( \int_0^{+\infty} du \left [ \psi_0 - \psi_0(+\infty) \right ] \right )
\int_{-\infty}^{+\infty} du I(u) \right \}, \nonumber
\end{eqnarray}
where $I(u) = \int_u^{u+a_x} du' e^{iqu'} f_{p'_k}^*(u') / a_x$ with $k=0, 1, 3$ 
for orientations $\theta=\pi/2$ (for ${\bm q}_{31}$), $\pi/6$ (for ${\bm q}_{21}$),
and $-\pi/6$ (for ${\bm q}_{23}$) respectively. Also,
\begin{widetext}
\begin{eqnarray}
& p_A(\theta=0) = \int_{-\infty}^{+\infty} du ~ e^{iqu} 
\left ( A_1^0 \partial_u f_{p_1}^* + {A_2^0}^* \partial_u f_{p_2} 
+ A_3^0 \partial_u f_{p_3}^* \right ),& \, (q=q_y, \text{ for direction } {\bm q}_2)
\label{eq:pA1}\\
& p_A(\theta=\pi/3) = \int_{-\infty}^{+\infty} du ~ e^{iqu} 
\left ( A_1^0 \partial_u f_{p_{11}}^* + {A_2^0}^* \partial_u f_{p_1}^* 
+ {A_3^0}^* \partial_u f_{p_0}^* \right ),& \, (q=q_y, \text{ for direction } -{\bm q}_1)
\label{eq:pA2}\\
& p_A(\theta=-\pi/3) = \int_{-\infty}^{+\infty} du ~ e^{iqu} 
\left ( {A_1^0}^* \partial_u f_{p_0} + {A_2^0}^* \partial_u f_{p_3}^* 
+ A_3^0 \partial_u f_{p_{33}}^* \right ),& \, (q=q_y, \text{ for direction } -{\bm q}_3)
\label{eq:pA3}\\
& p_A(\theta=\pi/2) = \int_{-\infty}^{+\infty} du ~ e^{iqu} 
\left ( A_1^0 \partial_u f_{p_{13}}^* + A_2^0 \partial_u f_{p_{33}} + {A_2^0}^* \partial_u f_{p_{11}}^*
+ {A_3^0}^* \partial_u f_{p_{31}} \right ),& \, (q=2q_x, \text{ for direction } {\bm q}_{31})
\label{eq:pA4}\\
& p_A(\theta=\pi/6) = \int_{-\infty}^{+\infty} du ~ e^{iqu} 
\left ( A_1^0 \partial_u f_{p_{12}}^* + {A_2^0}^* \partial_u f_{p_{21}} + A_3^0 \partial_u f_{p_2}
+ {A_3^0}^* \partial_u f_{p_{11}}^* \right ),& \, (q=2q_x, \text{ for direction } {\bm q}_{21})
\label{eq:pA5}\\
& p_A(\theta=-\pi/6) = \int_{-\infty}^{+\infty} du ~ e^{iqu} 
\left ( A_1^0 \partial_u f_{p_2} + {A_1^0}^* \partial_u f_{p_{33}}^* + {A_2^0}^* \partial_u f_{p_{23}}
+ A_3^0 \partial_u f_{p_{32}}^* \right ),& \, (q=2q_x, \text{ for direction } {\bm q}_{23})
\label{eq:pA6}
\end{eqnarray}
\end{widetext}

Given the condition $q_x=\sqrt{3}q_y/2$, from Eqs. (\ref{eq:gamma})--(\ref{eq:pA6})
and also Eqs. (\ref{eq:Ajpsi0})--(\ref{eq:Gj0}) it can be shown that these interfacial 
quantities $\gamma$, $\mu_k$, and $p_0$ are periodic functions of orientation
angle $\theta$ with a periodicity of $\pi/3$, consistent with the triangular
symmetry of the system. Thus results for directions ${\bm q}_2$ and $-{\bm q}_{1,3}$
(with $\theta=0,\pm \pi/3$) are equivalent; so are the results for directions
${\bm q}_{31}$, ${\bm q}_{21}$ and ${\bm q}_{23}$ (with $\theta=\pi/2,\pm \pi/6$).

In addition, the continuity condition at the solid-liquid interface (i.e., 
$u=0$) is given by
\begin{equation}
v_n \Delta \psi_0 = (1-m^2) \left [ \left ( {\bm \nabla}\delta\mu_{\psi} 
\right )_{\rm solid} - \left ( {\bm \nabla}\delta\mu_{\psi} \right )_{\rm liquid}
\right ] \cdot \hat{\bm n}.
\label{eq:continuity}
\end{equation}
Here $\delta\mu_{\psi}(u,s)$ is determined by the solutions of variations 
$\delta A_j = A_j' - A_j^0(\pm \infty)$ and $\delta \psi = \psi - \psi_0(\pm \infty)$
that are governed by
\begin{equation}
\left. \frac{\partial f}{\partial A_j'^*} \right |_1 = 0, \,
\frac{\partial \delta \psi}{\partial t} = (1-m^2) \nabla^2 \delta \mu_{\psi}
= (1-m^2) \nabla^2 \left. \frac{\partial f}{\partial \psi} \right |_1,
\label{eq:dAj_psi}
\end{equation}
where ``$|_1$'' refers to the expansion of $\partial f / \partial A_j'^*$
or $\partial f / \partial \psi$ to 1st order of $\delta A_j$ and $\delta \psi$.

Note that three key features have been intrinsically
incorporated in the above formulation of interfacial properties: 
i) meso-meso and meso-micro scale couplings, ii) crystalline anisotropy, 
and importantly, iii) compositionally generated elastic effects.
These will be further illustrated in the numerical results summarized in
the next section.

\section{Properties of alloy solid-liquid interface}
\label{sec:properties}

\begin{figure}
\centerline{\includegraphics[clip,width=0.48\textwidth]{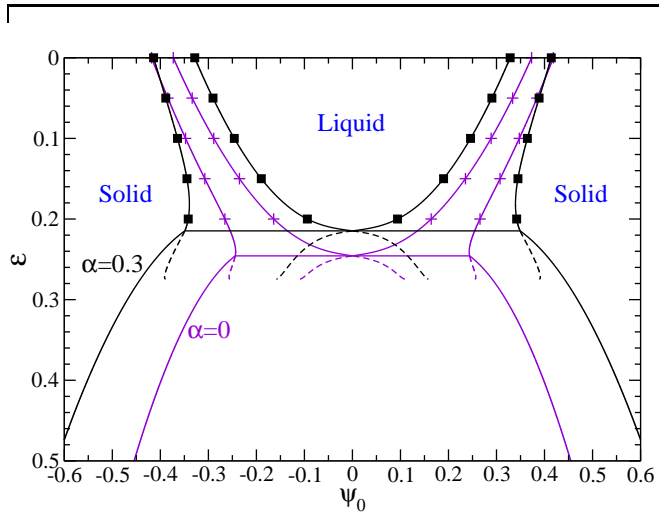}}
\caption{(Color online) Eutectic phase diagrams obtained from the complex 
amplitude model with 2D triangular symmetry, for $\alpha=0$ and $0.3$,
and $(n_0,w_0,g,g_2,u_0,K_0,v_1)=(-0.2,0.1,-1.8,-0.6,4,1,0)$.
Dashed lines are metastable extensions of the liquidus and solidus curves.
The symbols indicate some of the data points used in our calculations of 
solid-liquid interfacial properties. Phase diagrams for other values of $\alpha$
and the corresponding data points used are similar.}
\label{fig:eutectic}
\end{figure}

The analytic results given in Eqs. (\ref{eq:gamma})--(\ref{eq:pA6}) allow 
us to accurately and systematically determine the crystal-melt interfacial 
properties for binary alloys. Here we focus on a sample eutectic system, 
with model parameters $(n_0,w_0,g,g_2,u_0,K_0,v_1)=(-0.2,0.1,-1.8,-0.6,4,1,0)$,
and numerically calculate the interfacial free energy $\gamma$, kinetic 
coefficient $\mu_k$, and lattice pinning strength $p_0$ over various ranges 
of parameters $\alpha$ (the solute expansion coefficient) and $\epsilon$ 
(the effective reduced temperature) at 
different interface orientations $\theta$. The corresponding eutectic phase 
diagrams can be constructed analytically based on the free energy density 
determined by Eq. (\ref{eq:f}), with some sample results shown in 
Fig. \ref{fig:eutectic}. Although in these phase diagrams the $\psi_{s(l)}$ 
values for positive and negative branches of solidus (liquidus) lines are 
symmetric, the associated interfacial properties (i.e., $\gamma$, $\mu_k$, 
and $p_0$) are different due to the effect of compositional strain at the interface.
The corresponding numerical results are given below in Figs. \ref{fig:sig_eps_dpsi}%
--\ref{fig:p0_width}.

\subsection{Interfacial free energy}

\begin{figure}
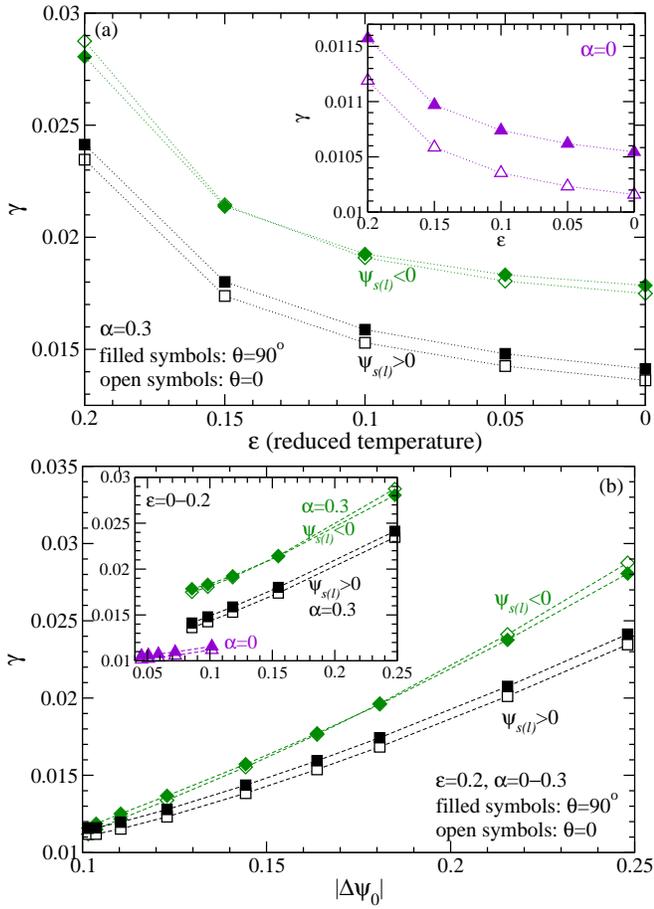

\centerline{\includegraphics[clip,width=0.48\textwidth]{bampl2048w01eps_al03al0_m0n_n02_sig.eps}}
\centerline{\includegraphics[clip,width=0.48\textwidth]{bampl2048w01eps_al_m0n_n02_sig_dpsi.eps}}
\caption{(Color online) (a) The solid-liquid interfacial free energy $\gamma$ as a function 
of reduced temperature $\epsilon$, for $\alpha=0.3$ and $\alpha=0$ (inset). 
(b) $\gamma$ as a function of miscibility gap $\Delta \psi_0$.
Filled or open symbols correspond to the $\hat{x}$ ($\theta=\pi/2$) or $\hat{y}$ 
($\theta=0$) interface orientation. Results for both positive and negative solidus 
(liquidus) alloy compositions $\psi_{s(l)}$ are shown.}
\label{fig:sig_eps_dpsi}
\end{figure}

\begin{figure}
\centerline{\includegraphics[clip,width=0.5\textwidth]{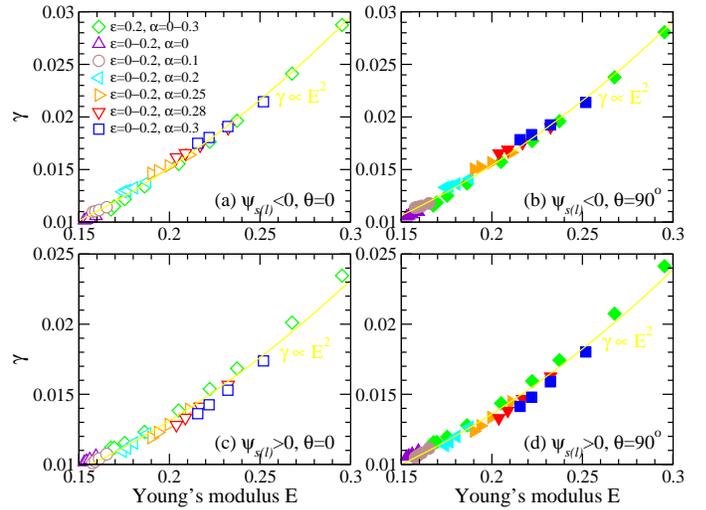}}
\caption{(Color online) Scaling of $\gamma$ as a function of Young's modulus $E$.}
\label{fig:sig_E}
\end{figure}

We have calculated the interfacial free energy $\gamma$ from Eq. (\ref{eq:gamma}) 
for various interface orientations, with $\theta$ values ranging from 0 to $\pi/3$ 
that are determined from directions of $k {\bm q}_i - l {\bm q}_j$ (with $i,j=1,2,3$ 
and $k,l$ integers). Our results shown in Fig. \ref{fig:sig_eps_dpsi} indicate 
that $\gamma$ increases with decreasing (or increasing) system temperature (or 
$\epsilon$ value) and increasing miscibility gap $\Delta \psi_0$, for all 
different values of $\alpha$ and $\epsilon$ parameters.
This is consistent with experimental measurements of e.g., Zn-Sn, 
Zn-In, and Al-Sn eutectic systems \cite{Eustathopoulos83}
and also the phase field modeling of Ni-Cu isomorphous alloy \cite{Wheeler93}.
This can be attributed to the larger composition gradient $\partial_u \psi_0^0$ 
around the interface for larger $\Delta \psi_0$. 
It leads to the increase of compositional free energy (see Eq. (\ref{eq:F})) 
which is absent in the excess configurational entropy theory for 
single-component systems \cite{Spaepen75}.
Interestingly, at a given interface orientation results 
of $\gamma$ for different ranges of $\epsilon$ and $\alpha$ fall onto 
a scaling relation as a function of Young's modulus $E$, 
as illustrated in Fig. \ref{fig:sig_E}. 
Parameters of this scaling curve depend on 
the selection of either positive ($\psi_{s(l)}>0$) or negative ($\psi_{s(l)}<0$) 
branch of solidus-liquidus lines due to different effect of compositional 
strain caused by nonzero $\alpha$ (see below for more discussions). 
Actually similar type of data collapse vs $E$ has been obtained from 
measurement data of surface free energy for some pure metals and alloys 
(although with different scaling relation for those solid-vapor results) 
\cite{Murr75}, yielding the correlation between solid surface 
energy and mechanical property of materials. 

\begin{figure}
\centerline{\includegraphics[clip,width=0.5\textwidth]{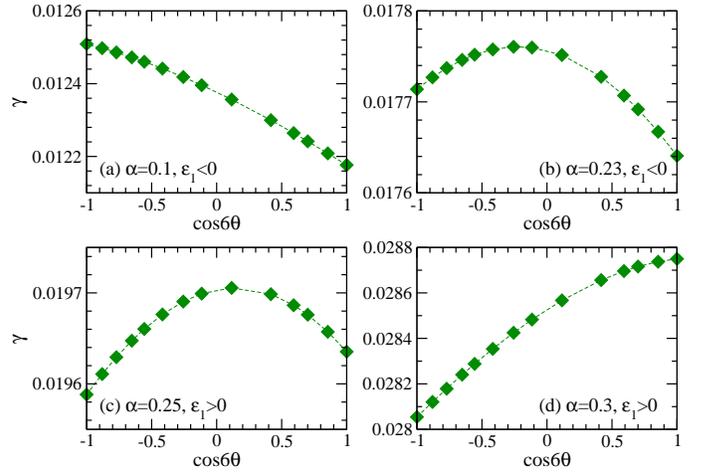}}
\caption{(Color online) Sample plots of interfacial free energy $\gamma$ vs $\cos 6\theta$, 
for $\psi_{s(l)}<0$, $\epsilon=0.2$, and $\alpha=0.1$, $0.23$, $0.25$, $0.3$.}
\label{fig:sig_cos6theta}
\end{figure}

\begin{figure}
\centerline{\includegraphics[clip,width=0.5\textwidth]{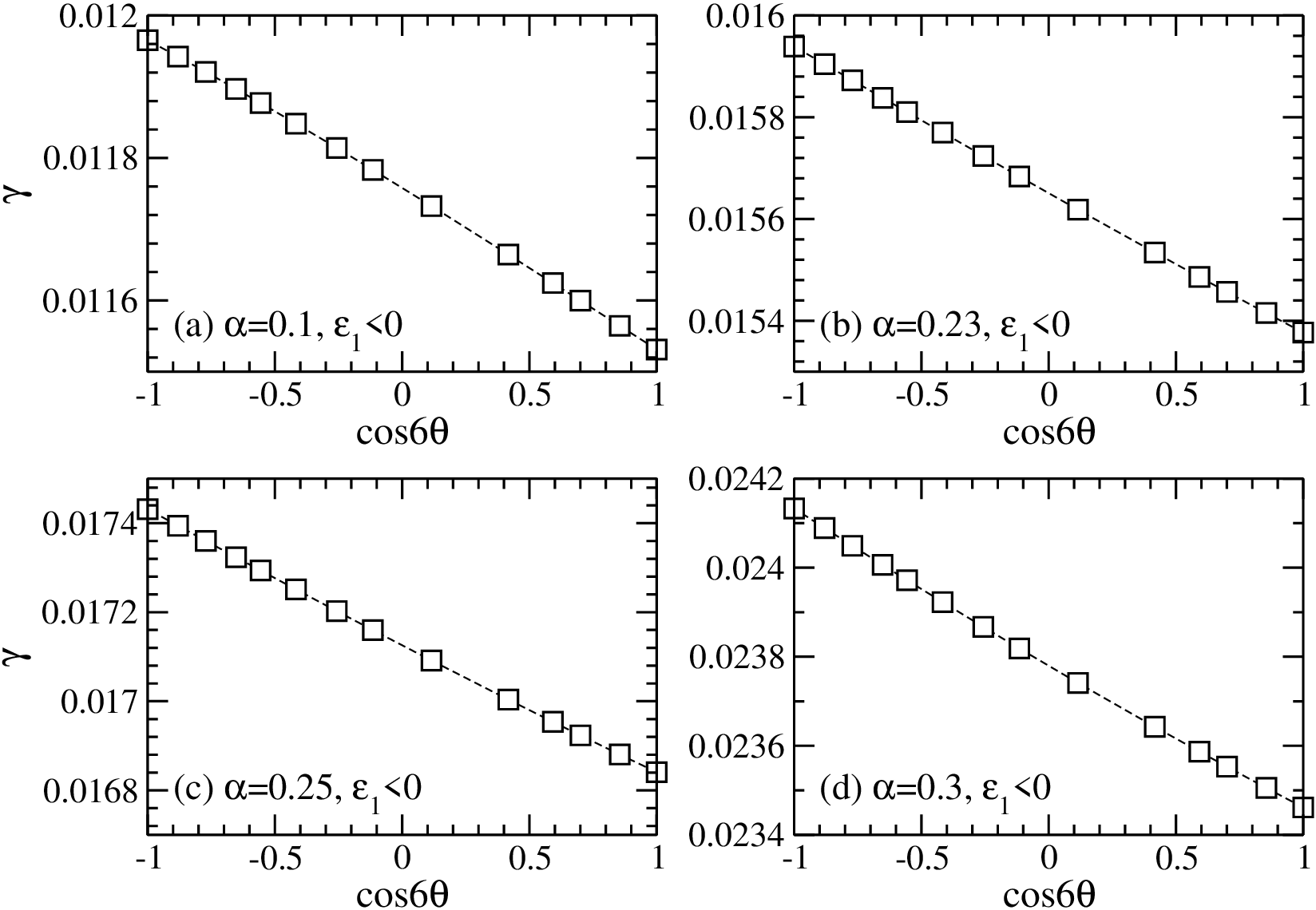}}
\caption{Sample plots of interfacial free energy $\gamma$ vs $\cos 6\theta$, 
for $\psi_{s(l)}>0$, $\epsilon=0.2$, and $\alpha=0.1$, $0.23$, $0.25$, $0.3$.}
\label{fig:sig+_cos6theta}
\end{figure}

\begin{figure}
\centerline{\includegraphics[clip,width=0.5\textwidth]{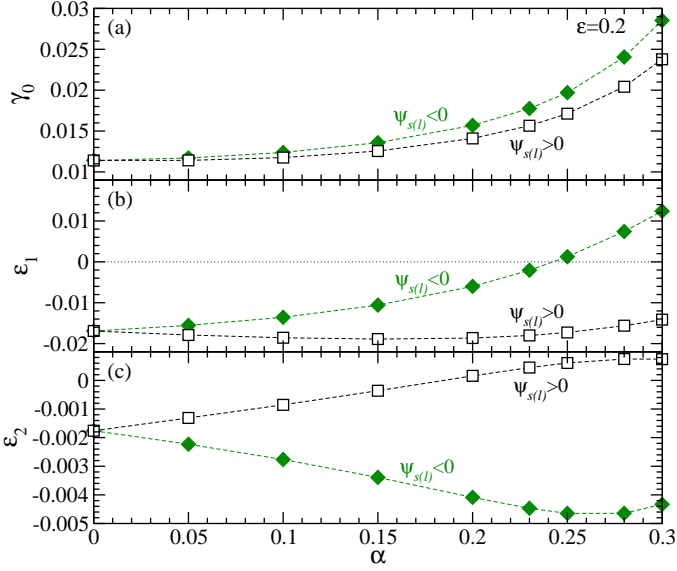}}
\caption{(Color online) $\gamma_0$ and anisotropy parameters $\varepsilon_1$, 
$\varepsilon_2$ vs $\alpha$ at $\epsilon=0.2$.}
\label{fig:sig_aniso}
\end{figure}

For triangular symmetry the anisotropy of interfacial energy can be 
represented by the expansion
\begin{equation}
\gamma = \gamma_0 (1 + \varepsilon_1 \cos 6\theta 
+ \varepsilon_2 \cos^2 6\theta + \cdots ),
\end{equation}
where $\varepsilon_1$ and $\varepsilon_2$ are anisotropic parameters.
In previous studies usually only the 1st-order expansion ($\varepsilon_1$)
is kept. The approach given in Eq. (\ref{eq:gamma}) can accurately determine 
even very weak anisotropy of $\gamma$, and our numerical data can be well 
fitted into this 2nd-order form, as shown in Figs. \ref{fig:sig_cos6theta} 
and \ref{fig:sig+_cos6theta} for negative ($\psi_{s(l)}<0$) and positive 
($\psi_{s(l)}>0$) solidus/liquidus branches respectively. 
The corresponding results for parameters 
$\gamma_0$, $\varepsilon_1$ and $\varepsilon_2$ are given in 
Fig. \ref{fig:sig_aniso}. Note that for $\psi_{s(l)}>0$ $\varepsilon_2$ is 
around an order of magnitude smaller than $\varepsilon_1$, but they can be 
of similar order for large enough $\alpha$ when $\psi_{s(l)}<0$.
Also $\gamma_0$ increases with the magnitude of $\alpha$ due to larger 
contribution of compositional strain. 

\begin{figure}
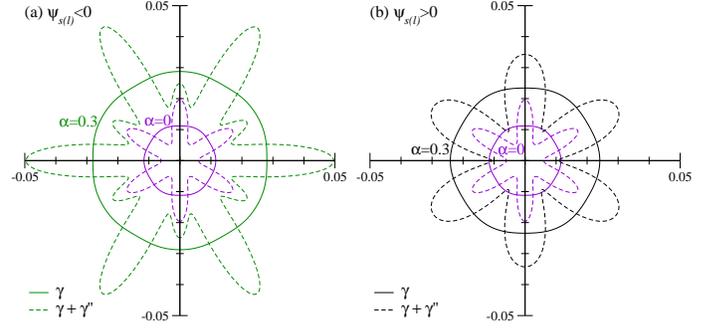

\centerline{\includegraphics[clip,width=0.25\textwidth]{bampl2048w01eps02al_m0n_n02_sig_d2sig_shape.eps}
\includegraphics[clip,width=0.25\textwidth]{bampl2048w01eps02al_m0n_n02_sig_d2sig+_shape.eps}}
\caption{(Color online) Polar plots of interfacial free energy $\gamma$ (solid lines) 
and stiffness $\gamma+\gamma''$ (dashed) for $\epsilon=0.2$, $\alpha=0$ and $0.3$.}
\label{fig:sig_shape}
\end{figure}

The important role played by the alloy compositional strain is further 
illustrated by its effect on the anisotropic parameter $\varepsilon_1$ 
(Fig. \ref{fig:sig_aniso}(b)): For $\psi_{s(l)}<0$ the increase of 
$\alpha$ leads to a reversal of sign of $\varepsilon_1$ (see also
Fig. \ref{fig:sig_cos6theta}), and thus 
a shape change (with a rotation of $30^{\circ}$) in the polar 
plots of $\gamma$ and interfacial stiffness $\gamma+\gamma''$ 
given in Fig. \ref{fig:sig_shape}(a); however no such changes 
occur for $\psi_{s(l)}>0$ (see Figs. \ref{fig:sig+_cos6theta}, 
\ref{fig:sig_aniso}(b), and \ref{fig:sig_shape}(b)), 
which instead gives a weak dependence of $\varepsilon_1$ on $\alpha$. 
This difference indicates an asymmetric effect of compressive vs 
tensile compositional stress at the interface.
For $\alpha>0$ (corresponding to larger size of atom A compared to atom B),
$\psi_{s(l)}<0$ indicates the abundance of smaller-size B atoms in the alloy.
In the solid surface layer where $\psi > \psi_s$  
(given $\psi_s < \psi_l <0$), a surface enrichment of larger atoms A occurs, 
resulting in a compressive solid surface layer with respect to the bulk, 
and its contribution to the excess interface free energy increases with 
$\alpha$ (see the interface elastic energy terms given in Eq. (\ref{eq:F_elastic})). 
When the anisotropy of this interface elastic energy contribution
is opposite to that of the non-compositional ones, the sign of 
$\varepsilon_1$ would then reverse at large enough $\alpha$ as 
seen in Fig. \ref{fig:sig_aniso}(b). On the other hand, our results of
$\varepsilon_1$ for $\psi_{s(l)}>0$ indicate that such anisotropy contrast 
does not exist (or is too weak) for the tensile-stress interface characterized
by the enrichment of smaller B atoms (at least for the range of $\alpha$ values
examined in our numerical calculations).

\begin{figure}
\centerline{\includegraphics[clip,width=0.5\textwidth]{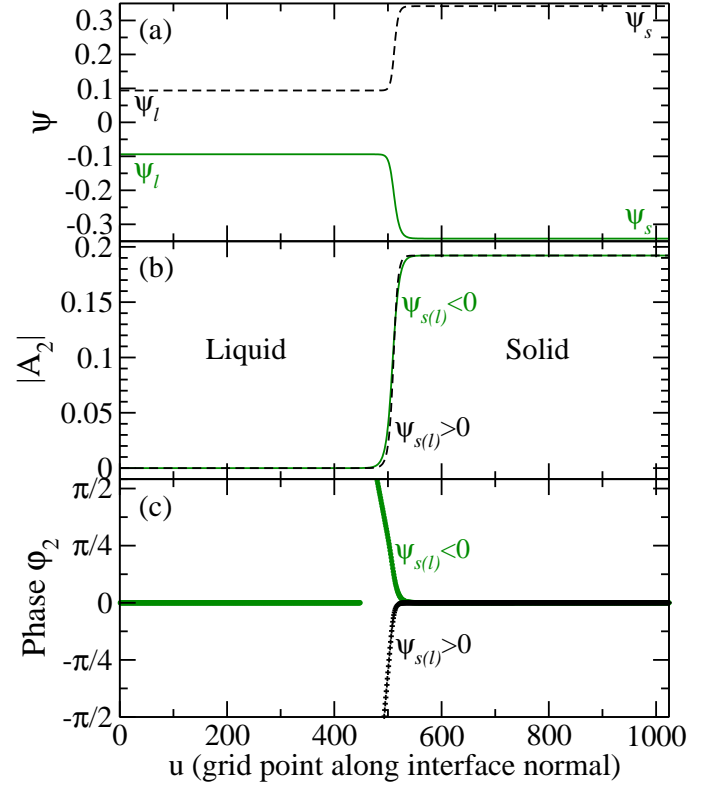}}
\caption{(Color online) Solid-liquid coexistent profiles of (a) $\psi$, (b) $|A_2|$, 
and (c) phase $\varphi_2$, for both positive and negative solidus/liquidus 
branches with $\epsilon=0.2$, $\alpha=0.3$, $\theta=0$, and
$(n_0,w_0,g,g_2,u_0,K_0,v_1)=(-0.2,0.1,-1.8,-0.6,4,1,0)$.}
\label{fig:A0psi}
\end{figure}

To verify this compositionally-induced interface stress, as in 
Sec. \ref{sec:ampl} we rewrite $A_j'=|A_j| \exp(i{\bm q}_j^0 \cdot {\bm u}_s)$, 
where ${\bm q}_j^0 \cdot {\bm u}_s \equiv \varphi_j$ is the phase of
the complex amplitude. Its gradient along the interface normal, 
$\partial_u \varphi_j$, will then yield the local strain of the system.
In the equilibrium state this gradient always vanishes in the bulk and 
would be nonzero around the solid-liquid interface if the above scenario 
of surface preferential segregation occurs. This has indeed been seen 
in our numerical results of $A_j^0$, as obtained from solving the 
1D amplitude equations (\ref{eq:Ajpsi0})--(\ref{eq:psi0}) at various 
interface orientations $\theta$.  A sample result is presented in 
Fig. \ref{fig:A0psi}, showing equilibrium interfacial profiles of 
amplitude and concentration for both $\psi_{s(l)}>0$ and $\psi_{s(l)}<0$ 
of solid-liquid coexistence. As shown in Fig. \ref{fig:A0psi}(c), 
the phase $\varphi_j=0$ in the solid bulk and is nonzero
only around the interface, yielding opposite sign of gradient 
$\partial_u \varphi_j$ for positive vs negative $\psi_{s(l)}$ when 
$\alpha \neq 0$. This gives rise to different type of interface strain,
i.e., tensile vs compressive, which is attributed to the phenomenon 
of interface segregation and deformation as discussed above. 
Similar results can be found in our calculations with other 
choices of parameters (e.g., $\epsilon$, $\theta$, and nonzero $\alpha$).
Note that the surface/interface stress identified here is different from
the single-component case, for which nonzero phase $\varphi_j$ and its 
spatial gradient around the interface have also been obtained in our
solutions of amplitude equations. However, in the alloy system studied 
here we have additional surface/interface stress generated by compositional 
effect, giving opposite type of strain for positive vs negative 
solidus/liquidus branch which is absent in the single-component system 
and only occurs when $\alpha \neq 0$.

\subsection{Interface kinetic coefficient}

\begin{figure}
\centerline{\includegraphics[clip,width=0.5\textwidth]{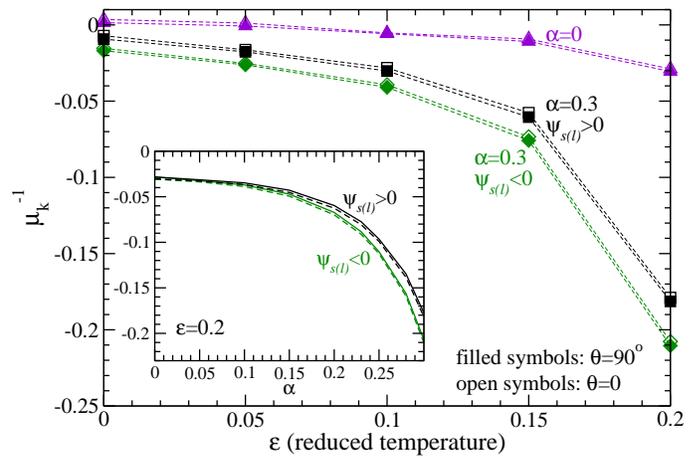}}
\caption{(Color online) Inverse kinetic coefficient $\mu_k^{-1}$ as a 
function of reduced temperature $\epsilon$, for $\alpha=0$ and $0.3$. 
Inset: $\mu_k^{-1}$ vs solute expansion coefficient $\alpha$ at $\epsilon=0.2$.}
\label{fig:zeta0_eps_al}
\end{figure}

For the kinetic coefficient $\mu_k$, to the best of our knowledge 
results for eutectic or isomorphous systems are still 
lacking, either from experiments or atomistic simulations, while only 
limited MD data is available for B2 and B33 ordered phases at
the melting temperature (Cu$_{50}$Zr$_{50}$ B2, Ni$_{50}$Al$_{50}$ B2 
and Ni$_{50}$Zr$_{50}$ B33 \cite{Kerrache08,Tang13,Wilson15}).
For the eutectic system examined here, our calculations
indicate a change of sign of $\mu_k$ from positive to negative 
at large enough miscibility gap (with large enough $\epsilon$, i.e., 
not close to the melting point, or large $\alpha$). This can be seen in 
Fig. \ref{fig:zeta0_eps_al}, which also shows the decrease of inverse kinetic 
coefficient $\mu_k^{-1}$ with the increase of $\epsilon$ and $\alpha$.
This is consistent with the previous phase field study 
\cite{re:elder01} which showed that for a growing interface with $\mu_k<0$, 
the relaxation of the interface profile would lag behind the advancing front. 
Note that these results are obtained under isothermal condition, with the
thermodynamic driving force $\Delta$ being the interface supersaturation 
of alloy concentration. It is different from many previous studies of 
single-component systems based on interface undercooling, although the basic 
mechanisms inside are similar \cite{re:langer80}.

\begin{figure}
\centerline{\includegraphics[clip,width=0.5\textwidth]{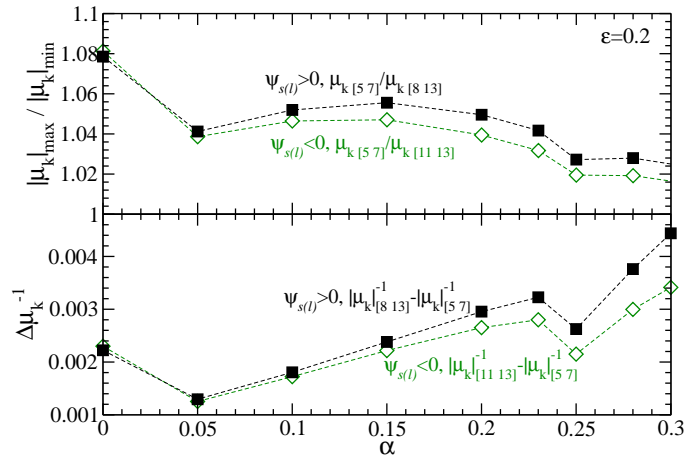}}
\caption{(Color online) Anisotropy of kinetic coefficient as a function of 
solute expansion coefficient $\alpha$ at $\epsilon=0.2$.}
\label{fig:dzeta0_alpha}
\end{figure}

Weak crystalline anisotropy of $\mu_k$ and its dependence on the compositional 
strain (or $\alpha$) are found in our calculations for this alloy system of 
triangular structure. For the sample results given in Fig.~\ref{fig:dzeta0_alpha}, 
around $2\%-8\%$ anisotropy (varying for different $\alpha$ values) is obtained 
at $\epsilon=0.2$, with minimum (maximum) magnitude of $\mu_k$ found at orientation 
$[8 ~13]$ with $\theta=7.59^{\circ}$ ($[5 ~7]$ with $\theta=13.90^{\circ}$) for 
$\psi_{s(l)}>0$, and at orientation $[11 ~13]$ with $\theta=21.79^{\circ}$ ($[5 ~7]$ with 
$\theta=13.90^{\circ}$) for $\psi_{s(l)}<0$.

\begin{figure}
\centerline{\includegraphics[clip,width=0.5\textwidth]{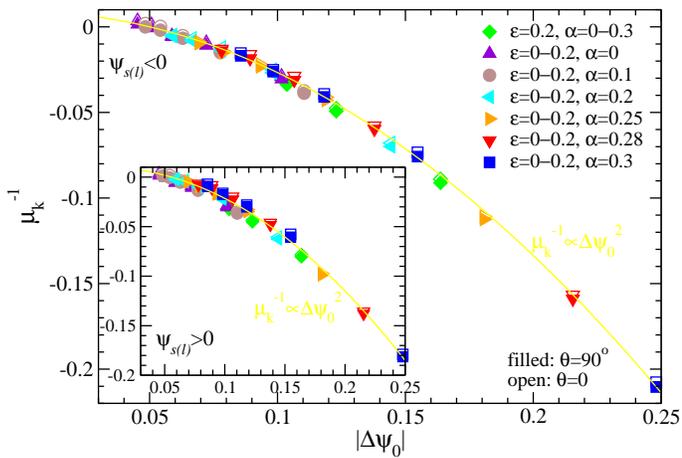}}
\caption{(Color online) Scaling of $\mu_k^{-1}$ as a function of miscibility gap 
$\Delta \psi_0$, for different ranges of $\epsilon$ and $\alpha$.}
\label{fig:zeta0_dpsi}
\end{figure}

More interestingly, a scaling behavior can be identified for $\mu_k$ 
when plotted against the miscibility gap $\Delta \psi_0$. 
As shown in Fig. \ref{fig:zeta0_dpsi}, data 
for various values of $\epsilon$ and $\alpha$ well converge to a universal 
curve $\mu_k^{-1} = a - b \Delta \psi_0^2$, although the scaling parameters
$a$ and $b$ are different for positive and negative solidus/liquidus branches.
Similar to the case discussed above, such a difference 
can be attributed to the effect of compositional strain. 
It causes the preferential segregation of larger (for $\psi_{s(l)}<0$) or 
smaller (for $\psi_{s(l)}>0$) atomic species on the solid surface and thus 
the compressive or tensile interface compositional stress, leading to 
different interface kinetics with smaller or larger values of $\mu_k$ 
respectively. This further demonstrates the important role played by the 
mesoscale coupling between the variation of alloy concentration field and 
the kinetics of interface structural profile.

Note that these results are for equal mobility of alloy constituents,
i.e., $m=0$. For nonzero mobility contrast
values of $\mu_k$ should be multiplied by a factor of $1-m^2$ 
(see Eq. (\ref{eq:mu_k})). This would then lead to $|\mu_k| \ll 1$ in the 
limit of $m \rightarrow \pm 1$ (with $M_A \gg M_B$ or $M_A \ll M_B$), 
consistent with the scenario of a frozen solid-liquid front due to the 
hindrance or pinning of one of the atomic components that is immobile.

\subsection{Lattice pinning strength}

\begin{figure}
\centerline{\includegraphics[clip,width=0.5\textwidth]{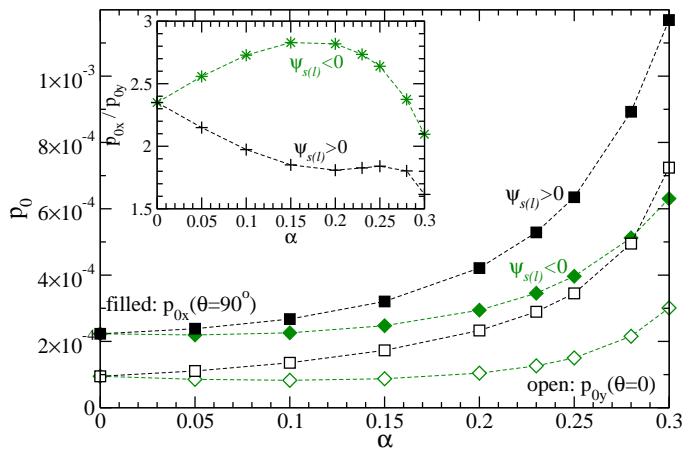}}
\caption{(Color online) Pinning strength $p_0$ as a function of solute expansion 
coefficient $\alpha$, for $\epsilon=0.2$ and two different interface orientations 
$\theta=0$, $\pi/2$. Inset: Anisotropy ratio of $p_0$ as a function of $\alpha$.}
\label{fig:p0_alpha}
\end{figure}

There exists a fundamentally different type of pinning effect that 
originates from the micro-meso length scale coupling between microscopic 
lattice structure and mesoscopic interface amplitudes and concentration,
as incorporated in Eqs. (\ref{eq:G-T}) and (\ref{eq:p0})--(\ref{eq:pA6}). 
This lattice coupling effect leads to two distinct modes of interface 
growth: As in the single-component case \cite{Huang13}, the solid 
front will advance in a continuous mode when the magnitude of the
thermodynamic driving force (i.e., $|\Delta|$) 
overcomes the lattice pinning strength $p_0$; otherwise when 
$|\Delta|<p_0$ the interface growth is characterized by a thermal 
activation and nucleation process, a scenario that is consistent with 
the crystal growth theory of Cahn \cite{re:cahn60,*re:cahn64}. For 
eutectic alloys our calculations show that $p_0$ is anisotropic 
as expected, as presented in Fig. \ref{fig:p0_alpha} which gives 
the results of $p_0$ calculated from Eqs. (\ref{eq:p0})--(\ref{eq:pA6}) 
for two interface growth directions $\theta=0$ ($\hat{y}$ direction with 
strength $p_{0y}$) and $\theta=\pi/2$ ($\hat{x}$ direction with strength $p_{0x}$).
We obtain large crystalline anisotropy of $p_0$, with ratio $p_{0x}/p_{0y}$
ranging from $1.61$ to $2.83$ (see the inset of Fig. \ref{fig:p0_alpha}).
Also for large enough compositional strain, this lattice pinning strength
increases with the magnitude of $\alpha$ for both $\psi_{s(l)}>0$ and $<0$
and at both interface orientations.

\begin{figure}
\centerline{\includegraphics[clip,width=0.5\textwidth]{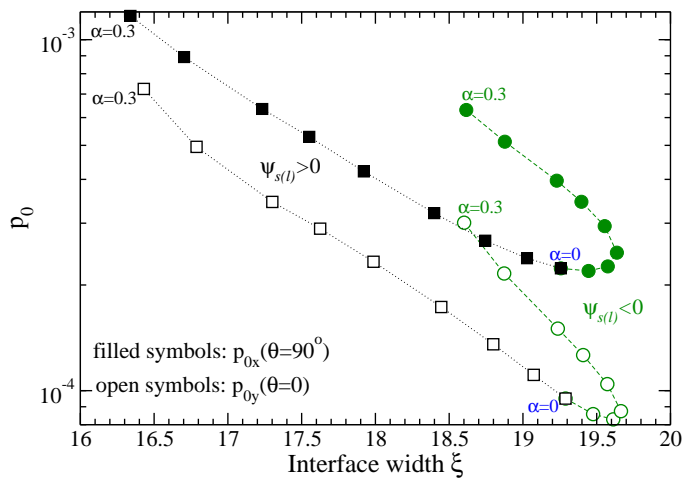}}
\caption{(Color online) $p_0$ as a function of interface thickness 
$\xi$, for $\epsilon=0.2$, $\alpha=0-0.3$,
$\psi_{s(l)}>0$ (squares) and $\psi_{s(l)}<0$ (circles).}
\label{fig:p0_width}
\end{figure}

Since this pinning effect is attributed to the nonadiabatic scale coupling 
at the interface, it is expected to increase with sharper interface and follow 
a universal relation $p_0 \sim \exp(-\alpha_p \xi)$ (with $\xi$ the interface
thickness and $\alpha_p$ a constant) for a given interface orientation 
as identified in pure systems \cite{Huang13}. 
However, for alloying systems this relation only holds within some limited
ranges of $\xi$, while for wide enough interfaces a hysteresis-type behavior 
of $p_0$ vs $\xi$ occurs, as shown in Fig. \ref{fig:p0_width}. This behavior
arises from the coupling between structural and concentration profiles,
causing another asymmetric effect of nonzero compositional strain. 
For liquid-solid interfaces of the same width $\xi$, larger (smaller) 
alloy components are enriched in the solid surface layer for $\psi_{s(l)}<0$ 
($>0$) and $\alpha>0$, leading to larger (smaller) pinning strength of the 
underlying interface lattice as illustrated in Fig.~\ref{fig:p0_width}.

\section{Summary}
\label{sec:summary}

We have systematically identified the effects of length-scale coupling and 
compositional stresses on key interfacial properties and their scaling 
behaviors for binary alloys, based on a complex PFC amplitude model and 
the corresponding coarse-graining scheme and sharp/thin-interface analysis.
The method developed here can be directly applied to other 2D and 3D 
systems of different crystalline symmetries (as incorporated in the PFC 
models via modes selection and coupling).
All of them can be reduced to effective 1D interfacial systems for
different orientations as described above, making the calculation 
much more efficient as compared to previous atomistic computation efforts 
conducted in full dimensions. Importantly, this approach has incorporated 
system elasticity, crystalline symmetry and anisotropy, and couplings 
between different length scales that are missing in conventional 
continuum approaches.

It is also important to note that although what we study here is a 
model system, it can be parameterized to match to specific materials 
(via fitting to MD results of liquid-state direct correlation function and 
solid-state density amplitude \cite{re:wu07,re:jaatinen09}, to first-principles 
calculations \cite{re:muralidharan10}, or to thermodynamics databases 
\cite{Provatas10b}). As such our method would provide a viable route 
for quantitatively determining key interfacial properties including
interfacial free energy, kinetic coefficient, and lattice pinning
that govern the material growth and solidification processes.
Quantitative results of these interfacial properties should then depend on 
the atomistic specifics of the material examined. Nevertheless, the modeling 
scheme presented above is based on general principles of symmetry and 
length scale couplings (micro-meso and meso-meso). Thus some results obtained 
here, in particular the scaling behaviors identified, are expected to be 
intrinsic and not sensitive to microscopic details of alloy constituents 
and their interactions, a feature that is important for gaining fundamental 
insights of material properties.

\begin{acknowledgments}
This work was supported by the National Science Foundation under Grant No. DMR-0845264.
\end{acknowledgments}

\appendix
\section{Derivation of interface equations of motion with lattice pinning}
\label{sec:dev_interface}

We conduct a coarse-graining analysis of the nonadiabatic amplitude equations 
(\ref{eq:A1})--(\ref{eq:psi}) and derive the corresponding anisotropic sharp/thin-interface 
equations of motion. Two derivation steps are needed: (i) The projection operator
method \cite{re:elder01} is used to obtain the interface equations in local
curvilinear coordinates around a certain interface orientation. 
(ii) A variation method similar to that of Refs. \cite{Herring51,re:karma98}
is applied to the free energy of the system, to identify the anisotropic 
form of the Gibbs-Thomson relation. Detailed results are given below,
including the general formulation of interface equations and the corresponding
interfacial quantities (in the case of varying $n_0$), as well as the simplified
case of $n_0=const.$ which leads to Eqs. (\ref{eq:G-T})--(\ref{eq:dAj_psi}) in
Sec. \ref{sec:interface_eqs}.

\subsection{Interface equations for a fixed orientation $\theta=\theta_0$}
\label{sec:dev_theta0}

Using the standard procedure of sharp/thin-interface approach 
\cite{re:elder01,re:karma98,Huang13}, we first examine separately the inner 
region close to the interface and the outer region far from it via expanding 
the variables in orders of a small parameter $\varepsilon$ (the interface 
P\'eclet number), and then match the inner and outer solutions to determine
the interfacial boundary conditions.
For a certain liquid-solid interface orientation $\theta_0$, we can assume 
$\theta(s) = \theta_0 + \varepsilon \tilde{\theta}$, where 
$\theta$ is the angle between the local normal direction $\hat{\bm n}$ of
the interface and the $\hat{y}$ axis. We also assume different scalings along 
and perpendicular to the interface normal in the inner region, i.e., $u/\xi$
and $\varepsilon s/\xi$ (with $\xi$ the interface thickness) for local curvilinear 
coordinates $u$ and $s$. At $\mathcal{O}(1)$ we obtain the 1D equilibrium
solutions $A_j'=A_j^0(u)$, $n_0=n_0^0(u)$, and $\psi=\psi_0(u)$ for a planar 
interface oriented at $\theta_0$ in the liquid-solid coexistence, i.e.,
\begin{equation}
\left. \frac{\delta \mathcal{F}}{\delta A_j'^*} \right |_0 = 0, \quad
\partial_u^2 \mu_{\psi}^0 = 0, \quad \partial_u^2 \mu_{n_0}^0 = 0,
\label{eq:Ajpsi0}
\end{equation}
where
\begin{eqnarray}
&& \left. \frac{\delta \mathcal{F}}{\delta A_j'^*} \right |_0 
= \left. \frac{\partial f}{\partial A_j'^*} \right |_0
+ \mathcal{G}_j^0 \left ( \mathcal{G}_j^0 - 2\delta_1^0 \right ) A_j^0 \nonumber\\
&& \qquad \qquad ~ -2q_0^2 \alpha \left [ \psi_0 \mathcal{G}_j^0 A_j^0
+ \mathcal{G}_j^0 (\psi_0 A_j^0) \right ], \label{eq:Aj0}\\
&& \mu_{\psi}^0 = \left. \frac{\delta \mathcal{F}}{\delta \psi} \right |_0
= \mu_{\psi}^{\rm eq} \nonumber\\
&& = \left. \frac{\partial f}{\partial \psi} \right |_0
-K_0 \partial_u^2 \psi_0 
- 2q_0^2 \alpha \sum_j \left ({A_j^0}^* \mathcal{G}_j^0 A_j^0 + {\rm c.c.} \right ),
\label{eq:psi0}\\
&& \mu_{n_0}^0 = \left. \frac{\partial f}{\partial n_0} \right |_0 = \mu_{n_0}^{\rm eq},
\label{eq:n00}
\end{eqnarray}
with
\begin{eqnarray}
&&\mathcal{G}_{1,3}^0 = \partial_u^2 + i \left ( \mp 2q_x \sin \theta_0 
- q_y \cos \theta_0 \right ) \partial_u, \nonumber\\
&&\mathcal{G}_2^0 = \partial_u^2 + 2i q_y \cos \theta_0 \partial_u.
\label{eq:Gj0}
\end{eqnarray}

These 1D 0th-order solutions $A_j^0$, $n_0^0$, and $\psi_0$ are used in the 
calculations at $\mathcal{O}(\varepsilon)$, which lead to the following 
interface equations after matching the inner and outer expansions. 
The continuity conditions at the solid-liquid interface are written as
\begin{eqnarray}
&v_n (\Delta n_0^0 - m \Delta \psi_0) = (1-m^2) \left [
\left. \partial_u \delta \mu_{n_0} \right |_{0^-}
- \left. \partial_u \delta \mu_{n_0} \right |_{0^+} \right ]& \nonumber\\
&= (1-m^2) \left [ \left ( {\bm \nabla}\delta\mu_{n_0} 
\right )_{\rm solid} - \left ( {\bm \nabla}\delta\mu_{n_0} \right )_{\rm liquid}
\right ] \cdot \hat{\bm n},& \label{eq:continuity1}\\
&v_n (\Delta \psi_0 - m \Delta n_0^0) = (1-m^2) \left [
\left. \partial_u \delta \mu_{\psi} \right |_{0^-}
- \left. \partial_u \delta \mu_{\psi} \right |_{0^+} \right ]& \nonumber\\
&= (1-m^2) \left [ \left ( {\bm \nabla}\delta\mu_{\psi} 
\right )_{\rm solid} - \left ( {\bm \nabla}\delta\mu_{\psi} \right )_{\rm liquid}
\right ] \cdot \hat{\bm n},& \label{eq:continuity2}
\end{eqnarray}
where $\Delta n_0^0 = n_0^0(+\infty) - n_0^0(-\infty)$, $\Delta \psi_0 = \psi_0(+\infty) 
- \psi_0(-\infty)$, $\delta\mu_{n_0} = \mu_{n_0} - \mu_{n_0}^{\rm eq}$, and
$\delta\mu_{\psi} = \mu_{\psi} - \mu_{\psi}^{\rm eq}$. The 1st-order outer 
equations governing the perturbations $\delta A_j = A_j' - A_j^0(\pm \infty)$,
$\delta n_0 = n_0 - n_0^0(\pm \infty)$, and $\delta \psi = \psi - \psi_0(\pm \infty)$
are given by
\begin{eqnarray}
&&\left. \frac{\partial f}{\partial A_j'^*} \right |_1 = 0, \label{eq:dAjn0psi}\\
&&\frac{\partial \delta n_0}{\partial t} 
= \nabla^2 \delta \mu_{n_0} + m\nabla^2 \delta \mu_{\psi}
= \nabla^2 \left. \frac{\partial f}{\partial n_0} \right |_1
+ m\nabla^2 \left. \frac{\partial f}{\partial \psi} \right |_1, \nonumber\\
&&\frac{\partial \delta \psi}{\partial t} 
= m\nabla^2 \delta \mu_{n_0} + \nabla^2 \delta \mu_{\psi}
= m\nabla^2 \left. \frac{\partial f}{\partial n_0} \right |_1
+ \nabla^2 \left. \frac{\partial f}{\partial \psi} \right |_1. \nonumber
\end{eqnarray}

At a moving interface the boundary condition is given by a generalized form
of the Gibbs-Thomson relation that incorporates the coupling and pinning of 
the underlying lattice structure, i.e., for an interface orientation $\theta_0$,
\begin{equation}
\mu_k^{-1} v_n  = -\Delta - \gamma \kappa -p_0 \sin(q h_n + \varphi) + \eta_v,
\label{eq:G-T0}
\end{equation}
where the thermodynamic driving force (interface supersaturation)
$\Delta = q_0^2 [\Delta \psi_0 \delta\mu_{\psi}(0,s) + \Delta n_0^0 \delta\mu_{n_0}(0,s)]$, 
$\gamma$ is the interfacial free energy expressed by Eq. (\ref{eq:gamma}) for a 
given $\theta=\theta_0$, and noise $\eta_v$ is governed by
\begin{equation}
\langle \eta_v \rangle = 0, \quad
\langle \eta_v(s,t) \eta_v(s',t') \rangle = 2D_v \delta(s-s') \delta(t-t'),
\end{equation}
with $D_v = \vartheta q_0^2 \Gamma k_BT \mu_k^{-1}$.
The kinetic coefficient $\mu_k$ is determined by
\begin{eqnarray}
&\mu_k^{-1} =& (1-m^2)^{-1} \int_{-\infty}^{+\infty} du 
\left \{ 2\sum\nolimits_j |\partial_u A_j^0|^2 \right. \nonumber\\
&& + q_0^2 \left [ \psi_0^2 - \psi_0^2(\pm \infty)
+ {n_0^0}^2 - {n_0^0}^2(\pm \infty) \right. \nonumber\\
&& \left. \left. -2m \left ( n_0^0\psi_0
- n_0^0 (\pm \infty)\psi_0(\pm \infty) \right ) \right ] \right \}.
\label{eq:mu_k_}
\end{eqnarray}
The lattice pinning strength $p_0$ and phase $\varphi$ can be written in a 
general form
\begin{equation}
p_0 e^{i\varphi} = 2i \left [ p_A(\theta_0) + p_{\psi}(\theta_0) 
+ p_{n_0}(\theta_0) \right ],
\end{equation}
where $p_{\psi}$ and $p_A$ are given in Eqs. (\ref{eq:p_psi})--(\ref{eq:pA6}).
Results of $p_{n_0}$ are similar to those of $p_{\psi}$; i.e., $p_{n_0}=0$ for 
${\bm q}_j$ orientations ($\theta_0=0, \pm \pi/3$), and $p_{n_0} \neq 0$ for
${\bm q}_{ij}$ orientations ($\theta_0=\pi/2,\pm \pi/6$), with
\begin{eqnarray}
&p_{n_0} = q_0^2& \left \{ \left [ \int_0^{+\infty} du n_0^0(u) \int_u^{+\infty} du'
\right. \right. \nonumber\\
&& \left. - \int_{-\infty}^0 du n_0^0(u) \int_{-\infty}^u du' \right ] I_0(u') \\
&& \left. - \left ( \int_0^{+\infty} du \left [ n_0^0 - n_0^0(+\infty) \right ] \right )
\int_{-\infty}^{+\infty} du I_0(u) \right \}, \nonumber
\end{eqnarray}
where $I_0(u) = \int_u^{u+a_x} du' e^{iqu'} f_{p_k}^*(u') / a_x$ with $k=0, 1, 3$ 
for ${\bm q}_{31}$ ($\theta_0=\pi/2$), ${\bm q}_{21}$ ($\theta_0=\pi/6$),
and ${\bm q}_{23}$ ($\theta_0=-\pi/6$) orientations.

\subsection{Variation method and anisotropic formulation}
\label{sec:variation}

For the case of varying local orientation $\theta$, we can simply replace $\theta_0$
by $\theta$ in the results given above, i.e., $\gamma(\theta_0) \rightarrow \gamma(\theta)$,
$\mu_k(\theta_0,m) \rightarrow \mu_k(\theta,m)$, and $p_0(\theta_0) \rightarrow p_0(\theta)$.
However, the corresponding anisotropic form of the Gibbs-Thomson relation is different,
with additional terms associated with gradients of surface/interface tension \cite{Herring51}.
Similar to the process of free energy variation used in Refs. \cite{Herring51,re:karma98}, 
for a system with non-moving ($v_n=0$) interface we have
\begin{equation}
\delta \left ( \mathcal{F} - \mathcal{F}_0 \right ) = 0,
\label{eq:var_F}
\end{equation}
given an infinitesimal perturbation of the interface with a perturbed hump 
around a reference point $(u=0,s=s_0)$. Here the system free energy 
$\mathcal{F} = \mathcal{F}_{\rm surface} + \mathcal{F}_{\rm bulk}$ with
$\mathcal{F}_{\rm surface} = \int ds \gamma(\theta)$ and 
$\mathcal{F}_{\rm bulk} = -PV + \int d{\bm r} (\mu_{\psi} \psi + \mu_{n_0} n_0)$ 
for a system of pressure $P$ and volume $V$, where $\mu_{\psi}$, $\psi$,
$\mu_{n_0}$, $n_0$ are determined from solutions of Eq. (\ref{eq:dAjn0psi}) 
in the outer region. $\mathcal{F}_0$ is the free energy of the equilibrium 
bulk state, i.e.,
\begin{eqnarray}
\mathcal{F}_0 & \simeq & \int ds \left \{ \int_0^{+\infty} du
\left [ \mu_{\psi}^{\rm eq} \psi_0(+\infty) + \mu_{n_0}^{\rm eq} n_0^0(+\infty) \right ]
\right. \nonumber\\
&& \left. + \int_{-\infty}^0 du 
\left [ \mu_{\psi}^{\rm eq} \psi_0(-\infty) + \mu_{n_0}^{\rm eq} n_0^0(-\infty) \right ] 
\right \} - PV, \nonumber
\end{eqnarray}
where we have assumed $d{\bm r} = \int ds \int du (1+u\kappa) \simeq \int ds \int du$
at the lowest order. Using the condition of Gibbs surface, we obtain
\begin{equation}
\mathcal{F} - \mathcal{F}_0 \simeq \int ds \gamma(\theta) 
+ \int ds \int_{-\infty}^{+\infty} du 
\left ( \delta \mu_{\psi} \psi + \delta \mu_{n_0} n_0 \right ). 
\label{eq:dF}
\end{equation}

It has been shown in Ref. \cite{Herring51} that 
\begin{equation}
\delta (\int ds \gamma(\theta))
= (\gamma + d^2 \gamma / d\theta^2) \kappa \delta V 
\equiv (\gamma + \gamma'') \kappa \delta V,
\end{equation}
where $\delta V = \int ds \delta u$. Also to first order of
perturbations, the variation of 2nd term in Eq. (\ref{eq:dF}) yields
\begin{eqnarray}
&&\delta \left [ \int ds \int du (\delta \mu_{\psi} \psi + \delta \mu_{n_0} n_0) 
\right ] \nonumber\\
&&\simeq \int ds \int du ~ \delta 
\left ( \delta \mu_{\psi} \psi + \delta \mu_{n_0} n_0 \right ) \nonumber\\
&&\simeq \left [ \delta \mu_{\psi}(0,s_0) \Delta \psi_0 
+ \delta \mu_{n_0}(0,s_0) \Delta n_0^0 \right ] \delta V,
\end{eqnarray} 
given $\delta \mu_{\psi}, \delta \mu_{n_0} \neq 0$ only around the interface $u=0$ and 
$\delta (\delta \mu_{\psi} \psi + \delta \mu_{n_0} n_0) 
= [\partial_u (\delta \mu_{\psi} \psi + \delta \mu_{n_0} n_0)] \delta u 
+ [\partial_s (\delta \mu_{\psi} \psi + \delta \mu_{n_0} n_0)] \delta s 
+ \mathcal{O}(\delta u^2, \delta s^2)$. Thus the variation of free energy 
in Eq. (\ref{eq:var_F}) becomes (for $q_0^2=1$ after rescaling)
\begin{equation}
-(\gamma + \gamma'') \kappa =  \delta \mu_{\psi}(0,s_0) \Delta \psi_0 
+ \delta \mu_{n_0}(0,s_0) \Delta n_0^0 \equiv \Delta.
\end{equation}
For the general case of moving interface with nonzero $v_n$ and the lattice pinning
effect given in Eq. (\ref{eq:G-T0}), the above relation can then be generalized
to an anisotropic from of the Gibbs-Thomson condition
\begin{eqnarray}
&\mu_k^{-1}(\theta,m) v_n  =& -\Delta - \left [ \gamma(\theta) + \gamma''(\theta) 
\right ] \kappa \nonumber\\
&& -p_0(\theta) \sin(q h_n + \varphi(\theta)) + \eta_v,
\label{eq:G-T_}
\end{eqnarray}
which leads to Eq. (\ref{eq:G-T}).

\subsection{Simplified case of $n_0=const.$}

Considering that for the liquid-solid interface of an alloy system the miscibility
gap is mostly determined by the concentration field $\psi$ and the variation of
$n_0$ is much smaller, for simplicity we can approximate $n_0$ as a constant. 
Applying $\partial n_0 / \partial t = 0$ to Eq. (\ref{eq:n0}) and combining it
with Eq. (\ref{eq:psi}), we then reduce the dynamic equation of $\psi$ to
\begin{eqnarray}
&&\partial \psi / \partial t 
= (1-m^2) \left \{ \nabla^2 \frac{\delta \mathcal{F}}{\delta \psi}
- \int_x^{x \pm a_x} \frac{dx'}{a_x} \int_y^{y+a_y} \frac{dy'}{a_y}
\right. \nonumber\\
&& \left. \left [ f_{p'_0} e^{i {\bm q}_{13} \cdot {\bm r}'}
+ f_{p'_1} e^{i {\bm q}_{12} \cdot {\bm r}'}
+ f_{p'_3} e^{i {\bm q}_{32} \cdot {\bm r}'} + {\rm c.c.} \right ] \right \}
+ {\bm \nabla} \cdot {\bm \eta}_{\psi_0}, \nonumber
\end{eqnarray}
where the noise term ${\bm \eta}_{\psi_0}$ is governed by
\begin{eqnarray}
& \langle {\bm \eta}_{\psi_0} \rangle = \langle {\bm \eta}_{\psi_0} \eta_j \rangle 
= \langle {\bm \eta}_{\psi_0} \eta_j^* \rangle = 0, & \nonumber\\
& \langle \eta_{\psi_0}^{\mu} \eta_{\psi_0}^{\nu} \rangle = 2(1-m^2) \vartheta_{\psi} 
\Gamma k_BT \delta ({\bm r} - {\bm r'}) \delta (t-t') \delta^{\mu\nu}.& \nonumber
\end{eqnarray}
Following the same procedure of sharp/thin-interface analysis described above
in Sec. \ref{sec:dev_theta0} and Sec. \ref{sec:variation}, we can simplify
the results of interface equations to those of Eqs. (\ref{eq:G-T})--(\ref{eq:dAj_psi}).

\bibliographystyle{apsrev4-1}
\bibliography{../references}

\end{document}